\documentclass[reprint,superscriptaddress,amsmath,amssymb,aps,prl,floatfix]{revtex4-2}

\usepackage{graphicx}
\usepackage{dcolumn}
\usepackage{bm}
\usepackage{hyperref}
\usepackage{xcolor}
\usepackage{float}
\usepackage{bigints}
\usepackage{nameref}

\newcommand{\calZ}{{\cal Z}}
\newcommand{\MeV}{\, {\rm MeV}}

\newcommand{\hmu}{\, {\hat{\mu}}}

\begin{document}

\title{Electric charge fluctuations from lattice QCD in the continuum limit}

\author{Szabolcs Bors\'anyi}
\affiliation{Department of Physics, Wuppertal University, Gaussstr.  20, D-42119, Wuppertal, Germany}

\author{Zolt\'an Fodor}
\affiliation{Pennsylvania State University, Department of Physics, State College, PA 16801, USA}
\affiliation{Pennsylvania State University, Institute for Computational and Data Sciences, University Park, PA 16802, USA}
\affiliation{Department of Physics, Wuppertal University, Gaussstr.  20, D-42119, Wuppertal, Germany}
\affiliation{Institute  for Theoretical Physics, ELTE E\"otv\"os Lor\' and University, P\'azm\'any P. s\'et\'any 1/A, H-1117 Budapest, Hungary}
\affiliation{J\"ulich Supercomputing Centre, Forschungszentrum J\"ulich, D-52425 J\"ulich, Germany}

\author{Jana N. Guenther}
\affiliation{Department of Physics, Wuppertal University, Gaussstr.  20, D-42119, Wuppertal, Germany}

\author{Paolo Parotto}
\affiliation{Dipartimento di Fisica, Universit\`a di Torino and INFN Torino, Via P. Giuria 1, I-10125 Torino, Italy}

\author{Attila P\'asztor}
\affiliation{Institute  for Theoretical Physics, ELTE E\"otv\"os Lor\' and University, P\'azm\'any P. s\'et\'any 1/A, H-1117 Budapest, Hungary}
\affiliation{MTA-ELTE Lendület "Momentum" Strongly Interacting Matter Research Group, Budapest, Hungary}

\author{Claudia Ratti}
\affiliation{ Department of Physics, University of Houston, Houston, TX 77204, USA}

\author{Volodymyr Vovchenko}
\affiliation{ Department of Physics, University of Houston, Houston, TX 77204, USA}

\author{Chik Him Wong}
\affiliation{Department of Physics, Wuppertal University, Gaussstr.  20, D-42119, Wuppertal, Germany}

\date{\today}

\begin{abstract}
Electric charge fluctuations $\chi_n^Q$ allow comparisons between theory and experiment, but are elusive on the lattice due to severe cutoff effects. We use a 4HEX action to obtain $\chi_2^Q$ and, for the first time ever, $\chi_4^Q$ in the continuum limit.
We find disagreement with the hadron resonance gas (HRG) model, which we cannot 
explain with finite volume effects. 
We include light meson interactions in the HRG model via the S-matrix, reducing 
the tension for $\chi_4^Q$, but worsening the agreement for $\chi_2^Q$.
We propose measuring the ratio $\chi_4^Q/\chi_2^Q$ at the LHC to investigate this tension.    
\end{abstract}

\maketitle

\paragraph{Introduction}

The phase structure of matter described by Quantum Chromodynamics (QCD) is very rich, and has been meticulously scrutinized in the last decades, especially in the low baryon density region. A low-temperature hadronic phase is set apart from a high-temperature quark gluon plasma (QGP) phase by a smooth crossover around $T=155-160$~MeV~\cite{Aoki:2006we,Bazavov:2018mes,Borsanyi:2020fev}. First-principles lattice simulations are the most robust tool to investigate QCD thermodynamics at low-to-moderate baryon densities in the nonperturbative regime, i.e. in the vicinity of the transition (see Refs.~\cite{Ratti:2018ksb,Guenther:2020jwe,Borsanyi:2025ttb} for a review).

Fluctuations of conserved charges of different orders enable a detailed study of the QCD medium, as they are sensitive to hadronic interactions~\cite{Vovchenko:2016rkn,Huovinen:2017ogf,Karthein:2021cmb}, remnants of possible critical behavior~\cite{Stephanov:2008qz,Stephanov:2011pb} and carry information about the effective degrees of freedom in the system~\cite{Jeon:2000wg,Asakawa:2000wh,Koch:2005vg,Cohen:2024ffx,Parra:2025fse,Gonzales:2026mfx}. Compared to other observables, fluctuations can be more directly related to measurable quantities in heavy-ion collisions (HICs), namely the event-by-event fluctuations in the number of particles carrying baryon number $B$, electric charge $Q$, or strangeness $S$~\cite{Bazavov:2012vg,Borsanyi:2013hza,Ratti:2018ksb,Borsanyi:2025kiv}, although a number of experimental effects ought to be accounted for before direct comparisons can be made~\cite{Vovchenko:2020tsr,koch2025exploringqcdphasediagram}. Fluctuations of the electric charge have a particular advantage in this regard: in contrast to fluctuations of baryon number and strangeness, they are measured directly and do not require the use of proxies to account for electrically neutral hadrons that are not measured~\cite{Kitazawa:2012at,Bellwied:2019pxh}.

From a theoretical standpoint, fluctuations of conserved charges are relatively easy to calculate in the grand canonical ensemble, and have been studied extensively with perturbative methods~\cite{Blaizot:2001vr,Andersen:2012wr,Haque:2013sja,Haque:2013qta}, with functional approaches~\cite{Fischer:2026vkc} such as the functional renormalization group (FRG)~\cite{Gao:2020qsj,Fu:2021oaw,Fu:2023lcm} and Dyson-Schwinger equations (DSE)~\cite{Isserstedt:2019pgx,Bernhardt:2022mnx,Lu:2025cls}, lattice simulations and models such as the hadron resonance gas (HRG)~\cite{Bazavov:2013dta,Vovchenko:2017xad,Alba:2017mqu,Bellwied:2019pxh,Vovchenko:2019pjl,Karthein:2021cmb} (see also Ref.~\cite{Vovchenko:2020lju} for a review). They are defined as derivatives of the free energy with respect to the corresponding chemical potentials:
\begin{equation*}
\chi_{ijk}^{BQS} = \frac{1}{VT^3} \frac{\partial^{i+j+k} \ln \calZ}{\partial \hmu_B^i \partial \hmu_Q^j \partial \hmu_S^k} \, \, ,
\end{equation*}
where we introduced the notation $\hmu_i = \mu_i/T$, which we will use hereafter. These quantities are also called susceptibilities, as they quantify how the system responds to a perturbation in the chemical potentials.

In recent years, many thermodynamic quantities of QCD have been calculated by more than one lattice collaboration, showing in general very good agreement between different discretizations of the QCD action: a fact that in turn further validates these results. This is the case for the QCD transition line~\cite{Bonati:2015bha,Bazavov:2018mes,Borsanyi:2020fev}, the equation of state  both at zero and finite chemical potential \cite{Borsanyi:2013bia,HotQCD:2014kol}, as well as low order fluctuations of conserved charges at zero~\cite{Borsanyi:2011sw,Bazavov:2012jq,Bellwied:2015lba,Bellwied:2019pxh,Bollweg:2021vqf,Goswami:2026hit} and finite density~\cite{Bollweg:2024epj,Borsanyi:2025kiv}. Fluctuations of the electric charge are a notable exception.

Electric charge fluctuations are particularly difficult observables in lattice QCD because they are largely dominated by pions, which in staggered discretizations suffer more than other hadrons from the breaking of taste symmetry at finite lattice spacing, due to their smaller mass. Although this effect disappears in the continuum theory, continuum extrapolations for electric charge fluctuations are especially delicate due to the need for very fine lattices. Continuum-extrapolated results exist from the HotQCD collaboration (with the HISQ action) for $\chi_2^Q$~\cite{Bollweg:2021vqf}, and from our collaboration (with the 4stout action) for $\chi_2^Q$~\cite{Borsanyi:2011sw,Bellwied:2015lba} and for $\chi_4^Q$~\footnote{The continuum extrapolation of $\chi_4^Q$ at $T=130$~MeV in Ref.~\cite{Bellwied:2015lba} used lattices with up to $N_\tau=32$. However, the $N_\tau=24$ and $N_\tau=32$ lattices had smaller volumes, which affected the continuum limit. Hence, we do not show this result in this work's plots.} (the latter only at $T=130$~MeV). Results from non-staggered discretizations also exist~\cite{Aarts:2014nba,Goswami:2026hit}, but only in exploratory studies and not continuum extrapolated.

In this Letter we present results for the second and -- for the first time in the literature -- fourth order electric charge fluctuations in the range $T=120-160 \MeV$ in the continuum limit.  We utilize our 4HEX action~\cite{Borsanyi:2023wno}, where the taste breaking is strongly suppressed (see Fig. 3 in the Supplemental Material of Ref.~\cite{Borsanyi:2023wno}) compared to the HISQ and 4stout actions. This allows us to perform a continuum limit without the need to resort to extremely fine lattices. 

We compare our results to predictions from the HRG model, finding no agreement for $\chi_4^Q$ and a tension on $\chi_2^Q$ below $T=130$~MeV. Since it has been suggested that the finite volume typical of lattice simulations can spoil the agreement with HRG model predictions~\cite{Bollweg:2021vqf}, we perform HRG calculations for the same finite volume we simulate on the lattice. We find that finite volume effects, although present, cannot account for the discrepancies we observe. We also perform dedicated lattice QCD simulations in smaller volumes for $T=130 \MeV$, and find that finite volume effects act in the opposite direction from that needed to ease the tension.

\begin{figure}
    \centering
    \includegraphics[width=\linewidth]{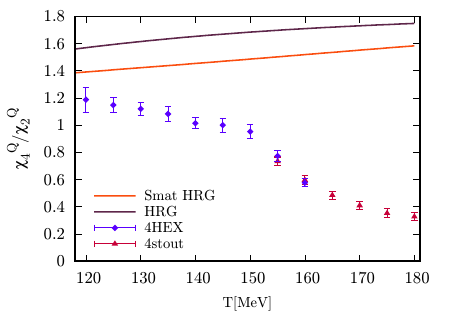}
    \caption{Continuum extrapolated results for the ratio $\chi_4^Q/\chi_2^Q$ with our 4HEX action (purple symbols), compared to results from the HRG model with (orange) and without (brown) the inclusion of light meson interactions with the S-matrix formalism (see Section \textit{S-matrix HRG model analysis}). We also include continuum-extrapolated results from our 4stout action at high temperature (red).
    }
    \label{fig:Q4Q2}
\end{figure}

We then carry out an analysis based on the S-matrix formulation of the HRG model~\cite{Dashen:1969ep}, including experimental input for hadron scattering in the $\pi$-$\pi$ and $\pi$-$K$ channels. We find that the tension between lattice and HRG results is reduced for $\chi_4^Q$, but at the same time, the agreement for $\chi_2^Q$ is lost. We show in Fig.~\ref{fig:Q4Q2} our continuum extrapolated results for the ratio $\chi_4^Q/\chi_2^Q$, compared to results from the HRG model with and without the inclusion of light meson scattering information. Although this inclusion reduces the tension, the results remain markedly different. We also include continuum-extrapolated results for the same ratio obtained with the 4stout action above $T=150$~MeV. We advocate measuring this ratio in HICs at the LHC to shed light on the role of meson interactions in the QCD medium, as existing measurements at the highest RHIC energy~\cite{Adamczyk:2014fia,Adare:2015aqk} do not have sufficient precision to do so.

\paragraph{Lattice analysis}\label{par:lattice}\def\parname{Lattice analysis}
In this Letter we employ our 4HEX action~\cite{Borsanyi:2023wno} with $N_f=2+1$ rooted staggered dynamical flavors with physical masses and the DBW2 action~\cite{Takaishi:1996xj} in the gauge sector, while in the fermion sector the links are improved with 4 levels of HEX smearing~\cite{Capitani:2006ni}. The main advantage of this action is the reduced taste breaking in the pion sector, particularly important for observables dominated by pion dynamics. We employ two different scale settings~\cite{Borsanyi:2023wno}, one based on the pion decay constant $f_\pi$, the other ($w_1$) on the modified Wilson-flow-based $w_0$~\cite{BMW:2012hcm}.

We perform simulations with temporal resolutions $N_\tau=8,10,12$~and~16, covering the range $T=120-165$~MeV with the $f_\pi$ scale setting (corresponding to $T=117-161$~MeV with the $w_1$ scale setting). In addition, we simulate an ensemble at $T_{f_\pi}=130$~MeV ($T_{w_1}=129$~MeV) on a $80^3\times20$ lattice to test the continuum extrapolation. The main ensembles were generated with aspect ratio $LT=4$ ($L\simeq 6~\mathrm{fm}$ at $T=130$~MeV), complemented with runs in smaller volumes $LT=2,3$ to assess sensitivity to the system size. The same action has already been used in Refs.~\cite{Borsanyi:2023wno,Borsanyi:2024xrx,Borsanyi:2025kiv}. The thermodynamics ensembles are listed in the Supplemental Material.

The measurements are performed following the standard practice~\cite{Allton:2002zi} based on stochastic sources where all statistical bias is canceled between disconnected contributions. Our implementation (see also Refs~\cite{Borsanyi:2013hza,Bellwied:2015lba,Borsanyi:2023wno}) benefits from the following improvements: low mode deflation with $\mathcal{O}(500)$ eigenvectors, truncated solver, hopping parameter expansion. These algorithmic ingredients are simpler versions of those used for the electric current correlators for the hadronic vacuum polarization~\cite{Borsanyi:2020mff}, and affect only the computation speed, not the result.  We used $2048$ stochastic sources in the truncated solver (used for the first derivative only) and additional $2048$ sources for the higher derivatives. The implementation was cross-checked on coarse ($16^3\times8$) lattices with the direct computation \cite{Borsanyi:2023wno} by means of the complete eigenvalue spectra (light and strange) in the reduced matrix formalism \cite{Hasenfratz:1991ax}.

\begin{figure}
    \centering
    \includegraphics[width=\linewidth]{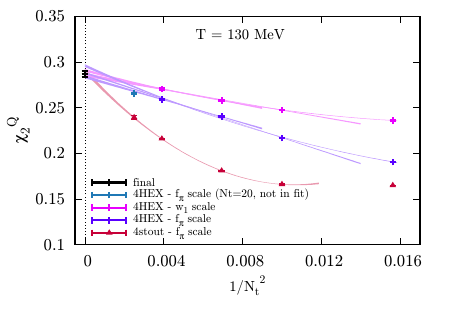}
    \includegraphics[width=\linewidth]{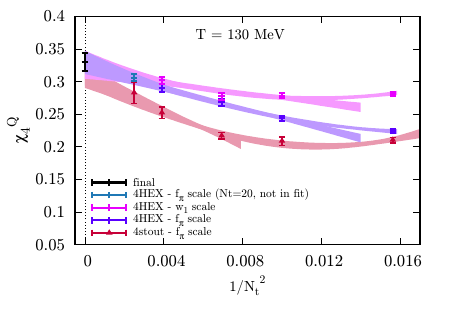}
    \caption{Example of continuum limits for the second (top) and fourth (bottom) electric charge susceptibilities at $T=130$~MeV. Several extrapolations to the continuum are shown, employing both scale settings in the 4HEX data (magenta and purple). We also show the result of simulations at $N_\tau=20$ employing the $f_\pi$ scale (teal), which is not included in the fits, but is found to be in agreement with them. The black points show the combined result from the different extrapolations (see main text for details). Additionally, we show the analogous limit with the 4stout action (red), where discretization errors are larger and thus finer lattices are needed.}
    \label{fig:QQ_QQQQ_T130_cont}
\end{figure}

\begin{figure}
    \centering
    \includegraphics[width=\linewidth]{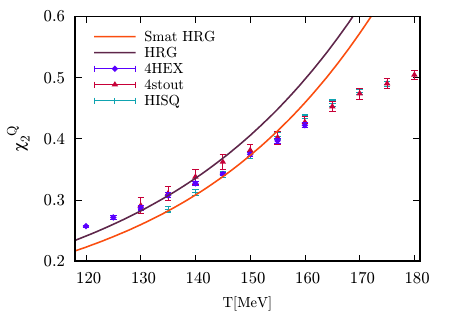}
    \includegraphics[width=\linewidth]{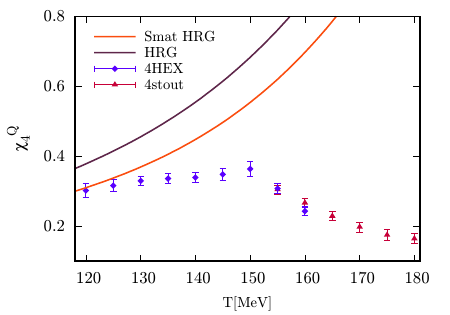}
    \caption{Continuum extrapolated results from the 4HEX action (purple) for the second (top) and fourth (bottom) electric charge susceptibilities as functions of the temperature. We compare with results from the HRG model without (brown) and with (orange) the inclusion of S-matrix treatment of $\pi$-$\pi$ and $\pi$-$K$ interactions, as well as with our 4stout~\cite{Bellwied:2015lba} (red) and HotQCD's HISQ~\cite{Bollweg:2021vqf} (cyan) actions, where available.}
    \label{fig:QQ_QQQQ_final}
\end{figure}

In Fig.~\ref{fig:QQ_QQQQ_T130_cont}, we illustrate how the continuum limit is approached with our 4stout action (red), for $\chi_2^Q$ at $T=130\MeV$, compared to the 4HEX results of this work (purple and magenta for the two scales). We first note that the latter have much smaller discretization effects: the same error we see at $N_\tau=20$ in 4stout is obtained at $N_\tau=12$ in 4HEX. Even more importantly, we see how the 4HEX action shows a much improved continuum scaling for the coarser lattice spacings, thus drastically reducing the systematic uncertainty stemming from the continuum extrapolation. It is clear from this that a continuum extrapolation from $N_\tau=8,10,12,16$ is feasible and more precise than its counterpart with our 4stout action. In Fig.~\ref{fig:QQ_QQQQ_T130_cont}, we also show a 4HEX result at $N_\tau=20$ with the $f_\pi$ scale setting to check the continuum extrapolation, finding good agreement. In order to estimate systematic uncertainties, we combine extrapolations with the two scale settings and perform the continuum limit with different ranges in $N_\tau$ and functional forms. We include linear fits to $N_\tau=10,12,16$ and parabolic fits to $N_\tau=8,10,12,16$, then combine them using the histogram method introduced in Ref.~\cite{Borsanyi:2020mff}. Only fits with a p-value above 0.01 are included. Given the excellent linear scaling of $\chi_2^Q$, in this case we also include one linear extrapolation from $N_\tau=12,16$ and a parabolic one from $N_\tau=10,12,16$ for each scale setting. The results of these extrapolations are shown in Fig.~\ref{fig:QQ_QQQQ_final}, compared to results from the ideal HRG model and the S-matrix HRG model (see next Section), as well as our previous results with the 4stout action (Ref.~\cite{Bellwied:2015rza} for $\chi_2^Q$) and from the HotQCD collaboration with the HISQ action~\cite{Bollweg:2021vqf}. As in Fig.~\ref{fig:Q4Q2}, for $\chi_4^Q$ we include a continuum extrapolation from the 4stout action starting with $T=155$~MeV. The 4stout dataset consists of the charge correlators on $32^3\times 8$, $40^3\times10$, $48^3\times12$, $64^3\times16$ and $80^3\times20$ lattices, extrapolated with linear and quadratic ans\"atze varying between two alternative scale settings ($f_\pi$ and $w_0$). Above the transition temperature the linear ansatz can describe the data. This dataset was already employed in Ref.~\cite{Abuali:2025tbd} to construct continuum estimates of all fluctuations of orders 2 and 4, including electric charge ones. However, for the purpose of that work, we imposed a smooth matching to the HRG model at low-$T$, which here we do not do. For the HRG calculations, we use the hadron list PDG2016+ from Ref.~\cite{Alba:2017mqu}, and checked our results not to be sensitive to this choice compared to other lists~\cite{SanMartin:2023zhv}. One can see that our results are in agreement with the HRG model for $\chi_2^Q$, except below $T=130$~MeV. This tension is visible due to the small uncertainties obtained from our analysis, but could not be observed previously because continuum extrapolations were not available at these temperatures. On the other hand, a large discrepancy exists for $\chi_4^Q$.

\begin{figure*}
    \centering
    \includegraphics[width=0.48\linewidth]{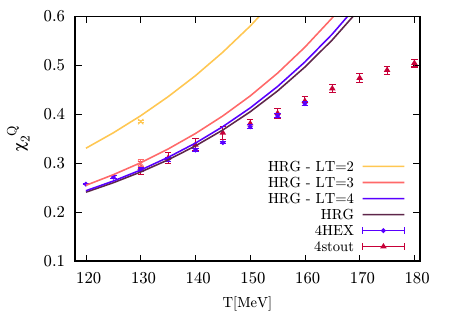}
    \includegraphics[width=0.48\linewidth]{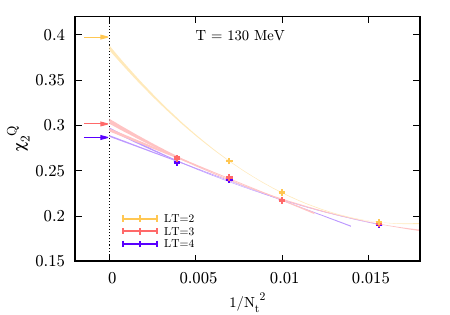}
    \includegraphics[width=0.48\linewidth]{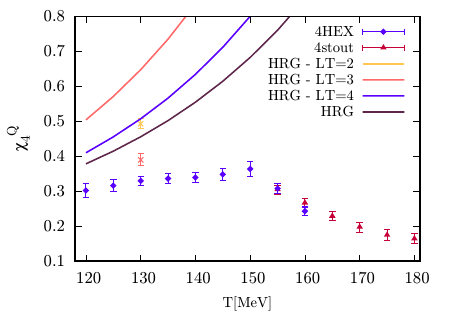}
    \includegraphics[width=0.48\linewidth]{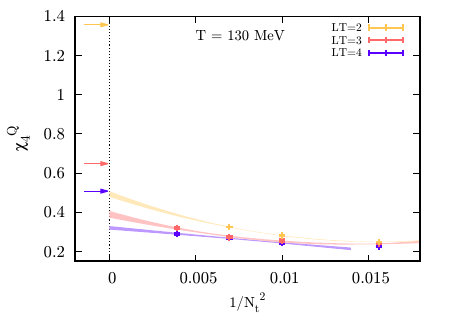}
    \caption{Left: second (top) and fourth (bottom) order electric charge fluctuations from the 4HEX action (purple) and the 4stout action (red), compared to the HRG model in the thermodynamic limit (brown) and with $LT=2,3,4$ (yellow, pink and purple). We see that a smaller volume brings the result up, and $LT=4$ is already close to the thermodynamic limit (very close in the case of $\chi_2^Q$). In the bottom plot, the yellow curve does not appear because it exceeds the plot range (see right panel). At $T=130$~MeV we also include the results of the continuum limits at $LT=3,2$ shown in the right panels. Right: continuum limit of $\chi_2^Q$ (top) and $\chi_4^Q$ (bottom) at $T=130$~MeV for our $LT=4$ (same as in Fig.~\ref{fig:QQ_QQQQ_T130_cont}), as well as $LT=3$ and $LT=2$ lattices. We note that finite volume effects are much larger for $\chi_4^Q$ than for $\chi_2^Q$.}
    \label{fig:QQ_volume}
\end{figure*}

\paragraph{Finite volume effects}
In this Section we investigate the role of finite volume effects, to determine whether the volume we employ ($LT=4$) is sufficiently close to the thermodynamic limit. In Fig.~\ref{fig:QQ_volume} we show HRG results for $\chi_2^Q$ (top left) and $\chi_4^Q$ (bottom left), both in an infinite and in a finite volume, compared to our new lattice results. In the right panels we show for $\chi_2^Q$ (top) and $\chi_4^Q$ (bottom) continuum extrapolations at $T=130$~MeV for volumes $LT=4,3,2$: HRG results at the corresponding volumes are shown as arrows. We see that a finite volume moves the curve upwards compared to the thermodynamic limit, for both observables and at all temperatures. This happens both in the HRG model and with our lattice results. We note that the effect on $\chi_2^Q$ at $LT=4$ is very small, of the order of $1-2\%$. Notably, finite volume effects for $\chi_2^Q$ in the HRG model and the lattice are comparable, while for $\chi_4^Q$ the HRG model shows much more pronounced effects. In any case, these effects go in the opposite direction as would be needed to ease the tension we observe.

\paragraph{S-matrix HRG model analysis}\label{par:Smat}\def\parname{S-matrix HRG model analysis}
Motivated by the observed tension between our lattice results and those from the HRG model, we investigate the possible role of non-resonant hadronic interactions on charge fluctuations, employing the S-matrix formulation of the HRG model and including the effect of $\pi$-$\pi$ and $\pi$-$K$ scattering. The correction to the free energy -- and consequently to fluctuations -- from the inclusion of an elastic scattering channel $s$ reads~\cite{Dashen:1969eg,Lo:2017lym,Andronic:2018qqt,Honek:2026exd}:
\begin{align*}
    \Delta \ln \calZ_s &= \frac{d_s \eta_s}{2 \pi^2} \bigintssss_{m_{\rm th}}^\infty dm \frac{1}{\pi} \frac{d \delta_s}{d m} \sum_{n=1}^\infty \frac{\left( -\eta_s \right)^{n+1}}{n^2} e^{\frac{n \mu_s}{T}} K_2\left(\frac{n m}{T}\right) 
\end{align*}
where $d_s$, $\eta_s$, $\delta_s$ are the spin multiplicity, quantum statistics factor and scattering phase shifts of the channel, $\mu_s = \mu_B B_s + \mu_Q Q_s + \mu_S S_s$, while $m_{\rm th}$ is the threshold mass. In a minimalist approach, we focus on the effect of primarily non-resonant scattering channels which cannot be included as hadron states in the HRG model, namely the S- and D-waves with $I=2$ in the $\pi$-$\pi$ channel~\cite{Garcia-Martin:2011iqs}, and the S-waves with $I=3/2,1/2$ for the $\pi$-$K$ channel~\cite{Pelaez:2016klv}. Details on the implementation are included in the Supplemental Material. We include the $\pi$-$K$ $I=1/2$ channel because the corresponding $\kappa$ meson is not included in our hadron list. The dominant effect comes from the S-wave in the $I = 2$ $\pi$-$\pi$ channel, which is negative (repulsive), while the contributions from other channels are small. We note that the contribution of $I = 2$ $\pi$-$\pi$ channel to isospin-averaged observables (such as the pressure) is largely canceled by the positive contribution from the S-wave $I = 0$ channel ($\sigma$ meson), which is why the $\sigma$ meson is usually excluded from thermal models~\cite{Venugopalan:1992hy,Broniowski:2015oha}. However, this cancellation does not occur in charge susceptibilities, where the contribution of the neutral $\sigma$ ($I = 0$) meson vanishes. Figure~\ref{fig:QQ_QQQQ_final} shows that the S-matrix HRG model improves the agreement of HRG with our lattice results for $\chi_4^Q$~\footnote{We note that a similar treatment of light meson interactions in the S-matrix HRG model, with similar results, was presented at the YSTAR2016 conference~\url{https://www.jlab.org/conferences/YSTAR2016/talks/thursday/huovinen.pdf}, but we are not aware of follow-up published results.}. At the same time, the agreement for $\chi_2^Q$ is worsened. If the ratio $\chi_4^Q/\chi_2^Q$ is considered, as shown in Fig.~\ref{fig:Q4Q2}, the tension is reduced but not resolved. We note that HRG models are often extended to incorporate $\pi$-$N$ interactions and repulsive baryon-baryon interactions~\cite{Vovchenko:2017xad}. The former were discussed in the context of the proton yield in HICs and  the baryon-charge correlator~\cite{Vovchenko:2018fmh,Andronic:2018qqt,Honek:2026exd}, while the latter, incorporated via an excluded volume, primarily affect baryon number susceptibilities~\cite{Karthein:2021cmb}. Using the implementation of these effects in \texttt{Thermal-FIST}~\cite{Vovchenko:2019pjl}, we have verified that they have a negligible influence on $\chi_2^Q$ and $\chi_4^Q$ at temperatures $T \lesssim 150$~MeV.

These results indicate that the description of pion interactions within the HRG model is incomplete, even within its extended S-matrix formulation, which takes into account the role of two-body, but not three-body (or more) interactions. In the Supplemental Material we further explore this issue by considering an alternative implementation of pion interactions based on an effective mass model (EMM)~\cite{Savchuk:2020yxc,Vovchenko:2020crk}. The EMM leads to a qualitatively different hierarchy of interaction effects where $\chi_2^Q$ is affected only weakly, while $\chi_4^Q$ receives a visible correction. This illustrates that the simultaneous description of $\chi_2^Q$ and $\chi_4^Q$ is sensitive to how pion interactions are modeled, and likely requires physics beyond the present two-body S-matrix HRG treatment. It is also worth noting that lattice studies of in-medium pion spectral properties above the chiral crossover find pion-like excitations with modified spectral structure~\cite{Lowdon:2022xcl}. Although these results apply at temperatures above the range considered here, they suggest that charge fluctuations may be sensitive to in-medium modifications of the pion sector beyond a two-body scattering treatment.

The $\chi_4^Q/\chi_2^Q$ ratio can potentially be accessed experimentally in HICs, where electric charge fluctuations are determined by event-by-event fluctuations in the numbers of charged particles. Although the measurements are not immune to experimental effects due to kinematic cuts, resonance decays, and (local) canonical effects, recent improvements allow one to systematically take those into account~\cite{Vovchenko:2024pvk,Parra:2025fse}. The discrepancy between our lattice results and HRG predictions is large and persists across all temperatures relevant to the freeze-out of charge fluctuations in HICs. Therefore, a measurement could test this experimentally and shed light on the reasons for this stark tension.

\paragraph{Conclusions}
In this Letter, we presented our continuum-extrapolated results for the second- and fourth-order electric charge fluctuations. Our results for $\chi_2^Q$ confirm our previous ones~\cite{Bellwied:2015rza}, while for $\chi_4^Q$ it is the first such result in the literature. We compared our results to the HRG model, finding fair agreement for $\chi_2^Q$ but a large tension for $\chi_4^Q$. We investigated the role of a finite volume both within the HRG, with calculations in a finite box, and within lattice QCD via dedicated runs at $T=130$~MeV on $LT=2,3$ lattices. We find that smaller volumes increase the values obtained for both observables, at all temperatures, both in the HRG model and from lattice simulations, but the effect at $LT=4$  is rather small. We then incorporated corrections to the HRG description via the S-matrix approach by including phase shifts for $\pi$-$\pi$ and $\pi$-$K$ channels. The inclusion of these interactions reduced the tension for $\chi_4^Q$, but at the same time spoiled the agreement for $\chi_2^Q$. The tension is also reduced for the ratio $\chi_4^Q/\chi_2^Q$, but remains sizable at all temperatures relevant for HICs. We thus suggest the experimental measurement of this quantity to clarify the source of this disagreement.

\paragraph{Acknowledgements}
This work was supported by the U.S. Department of Energy, Office of Science, Office of Nuclear Physics under contract numbers DE-SC0026065 and DE-SC0022023. Z. Fodor acknowledges funding from the DOE under the contract number DE-SC0025025. This material is based upon work supported by the National Science Foundation under grants No. PHY-2208724, PHY-2116686 and PHY-2514763, and within the framework of the MUSES collaboration, under Grant No. OAC-2103680. This material is also based upon work supported by the National Aeronautics and Space Agency (NASA) under Award Number 80NSSC24K0767. This work has also been supported by the Ministry of Culture and Science of the State of North Rhine-Westphalia, Germany under the funding code NW21-024-A. This research used resources of the Argonne Leadership Computing Facility, which is a DOE Office of Science User Facility supported under Contract DE-AC02-06CH11357. Access to computer time on Aurora (ALCF) was provided through the INCITE program. The authors gratefully acknowledge the Gauss Centre for Supercomputing e.V. (www.gauss-centre.eu) for funding the this project by providing computing time on the GCS Supercomputer Juwels/Booster at JSC, J\"ulich, Germany and Hazelhen/Hunter at HLRS, Stuttgart, Germany.

\section{Supplemental material}

\subsection{Treatment of pion interactions in the HRG model}

\begin{figure*}
    \centering
    \includegraphics[width=0.32\linewidth]{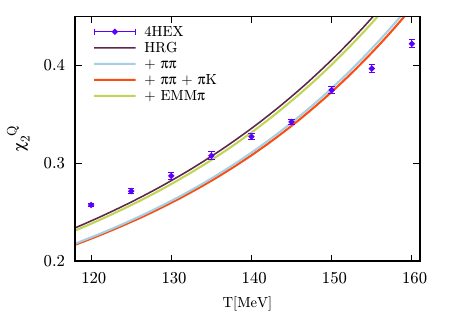}
    \includegraphics[width=0.32\linewidth]{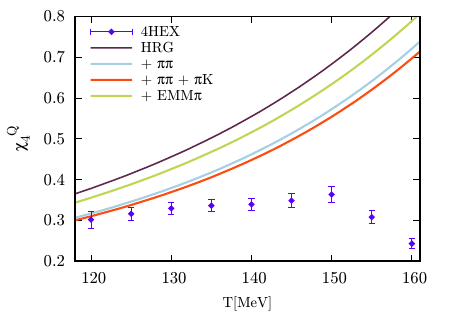}
    \includegraphics[width=0.32\linewidth]{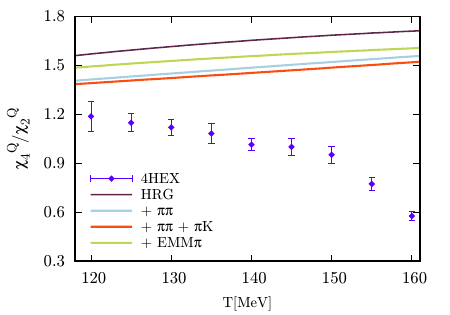}
    \caption{Electric charge susceptibilities $\chi_2^Q$ (left), $\chi_4^Q$ (middle), and their ratio $\chi_4^Q/\chi_2^Q$ (right). 
    Brown lines show the ideal HRG result, the light blue ones include the S-matrix correction from $\pi$-$\pi$ interactions, while the orange ones additionally include the $\pi$-$K$ contribution, as shown in the main text. 
    The green lines show the HRG result with pion interactions implemented through the effective mass model (EMM$\pi$). 
    Purple symbols denote the continuum-extrapolated lattice results obtained with the 4HEX action.
    }
    \label{fig:chi2Qchi4Q}
\end{figure*}

We explore how charge susceptibilities are affected by different implementations of interactions in the HRG model. Fig.~\ref{fig:chi2Qchi4Q} shows the effect of successively including $\pi$-$\pi$ interactions (light blue) and then $\pi$-$K$ interactions (orange, as in the main text). The comparison demonstrates that the dominant contribution comes from the $\pi$-$\pi$ channel, while the additional effect of the $\pi$-$K$ channel is relatively small.

The S-matrix corrections shown in Fig.~\ref{fig:chi2Qchi4Q} are evaluated in the Beth--Uhlenbeck form, using elastic phase shifts in the relevant non-resonant two-meson channels. For the $\pi$-$\pi$ contribution, we include the isotensor channels with $I=2$, namely the repulsive $J=0$ S-wave and the $J=2$ D-wave. The corresponding phase shifts are taken from the conformal parameterizations of Ref.~\cite{Garcia-Martin:2011iqs}. The $I=2$, $J=0$ channel uses the low- and intermediate-energy parameterizations matched at $\sqrt{s}=0.85$~GeV, while the $I=2$, $J=2$ channel is included up to $\sqrt{s}=1.42$~GeV. For the $\pi$-$K$ contribution, we include the S-wave channels with $I=3/2$ and $I=1/2$, using the conformal parameterizations of Ref.~\cite{Pelaez:2016klv}. The $I=3/2$ channel is repulsive and is integrated up to $\sqrt{s}=1.74$~GeV, while the attractive $I=1/2$ channel is restricted to the elastic region below the $K$-$\eta$ threshold. The appropriate spin degeneracy factors and charge weights are included for each isospin multiplet when computing $\chi_2^Q$ and $\chi_4^Q$.

As discussed above, the inclusion of pion interactions through the S-matrix approach affects both $\chi_2^Q$ and $\chi_4^Q$: it improves the agreement with the lattice results for the latter, but worsens it for the former. Since the effect on $\chi_2^Q$ is already sizable at temperatures as low as $T \simeq 120$~MeV, it is useful to explore alternative implementations of pion interactions, which may be sensitive, for example, to many-body effects not captured by the two-body S-matrix treatment.

As one such alternative, we consider the effective mass model (EMM)~\cite{Savchuk:2020yxc}, which was incorporated into the HRG model in Ref.~\cite{Vovchenko:2020crk} to model pion interactions at large isospin density. This is a quasiparticle model in which the free pion pressure is replaced by
\begin{align}
\label{eq:pEM}
p_\pi^{\rm EM}(T,\mu_\pi; m^*) = p_\pi^{\rm id}(T,\mu_\pi;m^*) + p_f(m^*) .
\end{align}
Here $\pi \in \{\pi^+,\pi^-,\pi^0\}$, $m^*$ is the effective pion mass, and $p_f(m^*)$ is a rearrangement term required for thermodynamic consistency. The effective mass is determined from the gap equation
\begin{align}
p_f'(m^*) = n_\sigma^{\rm id}(T,\mu_\pi;m^*) ,
\end{align}
where $n_\sigma^{\rm id}$ denotes the ideal-gas scalar density. The specific form of $p_f(m^*)$ defines the quasiparticle model. Here we use the form that reproduces the leading-order chiral perturbation theory result at $T=0$~\cite{Son:2000xc},
\begin{align}
\label{eq:chPTpf}
p_f(m^*) = \frac{(m^*)^2 f_\pi^2}{4}
\left[ 1 - \frac{m_\pi^2}{(m^*)^2} \right]^2 ,
\qquad f_\pi = 133~\text{MeV}.
\end{align}

The result of this calculation is shown in Fig.~\ref{fig:chi2Qchi4Q} by the green lines. The EMM has a negligible effect on $\chi_2^Q$, but produces a visible effect on $\chi_4^Q$. Thus, within this effective-mass implementation, interaction effects are more pronounced in higher-order susceptibilities. We note that the EMM framework of Ref.~\cite{Vovchenko:2020crk} was designed for large isospin density, where a single pion component dominates. At vanishing chemical potentials, this implementation appears to underestimate the overall effect of pion interactions and should be treated as a toy model rather than a systematic finite-temperature ChPT treatment of the pion gas at vanishing chemical potentials. Nevertheless, the fact that the EMM affects $\chi_4^Q$ much more strongly than $\chi_2^Q$, in contrast to the S-matrix treatment, may point toward a possible way of resolving the apparent difficulty of simultaneously describing $\chi_2^Q$ and $\chi_4^Q$ within HRG-based models.

\subsection{Finite temperature ensembles}\label{sec:ensembles}

In Tables \ref{tab:lt4stat}, \ref{tab:130stat} we show the number of configurations (separated by 10 RHMC updates) where we computed the fourth order cumulants with our 4HEX discretization. The dataset extends to $T=165$~MeV in the $f_\pi$-based scale setting. This analyses uses two scale settings (including the $w_1$ scale introduced in Ref.~\cite{Borsanyi:2023wno}). In the latter scale the (nominally) $T=165$~MeV ensembles have temperatures $T=161.0$, $161.5$ and $163.8$~MeV for $N_\tau=10,~12$ and 16, respectively. Thus, after interpolation the continuum limit is available up to $T=160$~MeV. The parametrization, tuning  and scale setting process is described in detail in Ref.~\cite{Borsanyi:2023wno}.

The 4stout dataset was first introduced in Ref.~\cite{Bellwied:2015lba} and was later extended in Refs.~\cite{Borsanyi:2020fev,Borsanyi:2021sxv}.

\begin{table}[h]
\begin{center}
\begin{tabular}{|c|c|c|c|c|}
\hline
$T$~[MeV] & $32^3\times 8$ & $40^3\times10$ & $48^3\times 12$ & $64^3\times 16$ \\
\hline
120& 17328& 22155& 14376& 11915\\
125& 13495& 21786& 14066& 11660\\
130& 16891& 21595& 13815& 11368\\
135& 16598& 21131& 13508& 11102\\
140& 16152& 20132& 12737& 10933\\
145& 16022& 19994& 12560& 10853\\
150& 15967& 20028& 12646& 10517\\
155& 16385& 21003& 13378& 11410\\
160& 16024& 23047& 15447& 12449\\
165& 18893& 25952& 18456& 10164\\
\hline
\end{tabular}
\end{center}
\caption{\label{tab:lt4stat}
List of 4HEX ensembles with the statistics of gauge configurations where the fourth order cumulants have been computed. This table shows the ensembles with aspect ratio $LT=4$, used in our main continuum extrapolation.
}
\end{table}

\begin{table}[h]
\begin{center}
\begin{tabular}{|c|c|}
\hline
lattice & statistics \\
\hline
$16^3\times8$ & 2132329 \\
$20^3\times10$ & 74626 \\
$24^3\times12$ & 68689 \\
$24^3\times8$ & 61025 \\
$30^3\times10$ & 10734 \\
$36^3\times12$ & 12233 \\
$48^3\times16$ & 10569 \\
$80^3\times20$ & 10209 \\
\hline
\end{tabular}
\end{center}
\caption{\label{tab:130stat}
Additional 4HEX configurations accumulated at $T=130$~MeV.
}
\end{table}

\bibliographystyle{unsrt}
\bibliography{thermo,paolo,volodya}

\begin{thebibliography}{10}

\bibitem{Aoki:2006we}
Y.~Aoki, G.~Endrodi, Z.~Fodor, S.D. Katz, and K.K. Szabo.
\newblock {The Order of the quantum chromodynamics transition predicted by the
  standard model of particle physics}.
\newblock {\em Nature}, 443:675--678, 2006.

\bibitem{Bazavov:2018mes}
A.~Bazavov et~al.
\newblock {Chiral crossover in QCD at zero and non-zero chemical potentials}.
\newblock {\em Physics Letters B}, 795:15--21, Aug 2019.

\bibitem{Borsanyi:2020fev}
Szabolcs Borsanyi, Zoltan Fodor, Jana~N. Guenther, Ruben Kara, Sandor~D. Katz,
  Paolo Parotto, Attila Pasztor, Claudia Ratti, and Kalman~K. Szabo.
\newblock {The QCD crossover at finite chemical potential from lattice
  simulations}.
\newblock {\em Phys. Rev. Lett.}, 125:052001, 2020.

\bibitem{Ratti:2018ksb}
Claudia Ratti.
\newblock {Lattice QCD and heavy ion collisions: a review of recent progress}.
\newblock {\em Rept. Prog. Phys.}, 81(8):084301, 2018.

\bibitem{Guenther:2020jwe}
Jana~N. Guenther.
\newblock {Overview of the QCD phase diagram: Recent progress from the
  lattice}.
\newblock {\em Eur. Phys. J. A}, 57(4):136, 2021.

\bibitem{Borsanyi:2025ttb}
Szabolcs Borsanyi and Paolo Parotto.
\newblock {\em {The QCD phase diagram}}, volume Encyclopedia of Particle
  Physics.
\newblock Elsevier, 2026.

\bibitem{Vovchenko:2016rkn}
Volodymyr Vovchenko, Mark~I. Gorenstein, and Horst Stoecker.
\newblock {van der Waals Interactions in Hadron Resonance Gas: From Nuclear
  Matter to Lattice QCD}.
\newblock {\em Phys. Rev. Lett.}, 118(18):182301, 2017.

\bibitem{Huovinen:2017ogf}
Pasi Huovinen and Peter Petreczky.
\newblock {Hadron Resonance Gas with Repulsive Interactions and Fluctuations of
  Conserved Charges}.
\newblock {\em Phys. Lett.}, B777:125--130, 2018.

\bibitem{Karthein:2021cmb}
Jamie~M. Karthein, Volker Koch, Claudia Ratti, and Volodymyr Vovchenko.
\newblock {Constraining the hadronic spectrum and repulsive interactions in a
  hadron resonance gas via fluctuations of conserved charges}.
\newblock {\em Phys. Rev. D}, 104(9):094009, 2021.

\bibitem{Stephanov:2008qz}
M.A. Stephanov.
\newblock {Non-Gaussian fluctuations near the QCD critical point}.
\newblock {\em Phys.Rev.Lett.}, 102:032301, 2009.

\bibitem{Stephanov:2011pb}
M.A. Stephanov.
\newblock {On the sign of kurtosis near the QCD critical point}.
\newblock {\em Phys.Rev.Lett.}, 107:052301, 2011.

\bibitem{Jeon:2000wg}
S.~Jeon and V.~Koch.
\newblock {Charged particle ratio fluctuation as a signal for QGP}.
\newblock {\em Phys.Rev.Lett.}, 85:2076--2079, 2000.

\bibitem{Asakawa:2000wh}
Masayuki Asakawa, Ulrich~W. Heinz, and Berndt Muller.
\newblock {Fluctuation probes of quark deconfinement}.
\newblock {\em Phys.Rev.Lett.}, 85:2072--2075, 2000.

\bibitem{Koch:2005vg}
V.~Koch, A.~Majumder, and J.~Randrup.
\newblock {Baryon-strangeness correlations: A Diagnostic of strongly
  interacting matter}.
\newblock {\em Phys. Rev. Lett.}, 95:182301, 2005.

\bibitem{Cohen:2024ffx}
T.~D. Cohen and L.~Ya. Glozman.
\newblock {On interpretation of fluctuations of conserved charges at high T}.
\newblock {\em Eur. Phys. J. A}, 60(8):170, 2024.

\bibitem{Parra:2025fse}
Jonathan Parra, Roman Poberezhniuk, Volker Koch, Claudia Ratti, and Volodymyr
  Vovchenko.
\newblock {Indications for Freeze-Out of Charge Fluctuations in the Quark-Gluon
  Plasma at the LHC}.
\newblock {\em Phys. Rev. Lett.}, 135(24):242302, 2025.

\bibitem{Gonzales:2026mfx}
Jonathan Gonzales, Alejandro Florez, Johannes Jahan, Angel R.~Nava Acuna, Naman
  Mehndiratta, and Claudia Ratti.
\newblock {Partial Pressure Contributions of Hadron Families to the QCD
  Equation of State}.
\newblock 6 2026.

\bibitem{Bazavov:2012vg}
A.~Bazavov, H.T. Ding, P.~Hegde, O.~Kaczmarek, F.~Karsch, et~al.
\newblock {Freeze-out Conditions in Heavy Ion Collisions from QCD
  Thermodynamics}.
\newblock {\em Phys.Rev.Lett.}, 109:192302, 2012.

\bibitem{Borsanyi:2013hza}
S.~Borsanyi, Z.~Fodor, S.D. Katz, S.~Krieg, C.~Ratti, et~al.
\newblock {Freeze-out parameters: lattice meets experiment}.
\newblock {\em Phys.Rev.Lett.}, 111:062005, 2013.

\bibitem{Borsanyi:2025kiv}
Szabolcs Bors{\'a}nyi, Zolt{\'a}n Fodor, Jana~N. Guenther, Piyush Kumar, Paolo
  Parotto, Attila P{\'a}sztor, and Chik~Him Wong.
\newblock {Finite density QCD phase structure from strangeness fluctuations}.
\newblock {\em Phys. Rev. D}, 113(5):054507, 2026.

\bibitem{Vovchenko:2020tsr}
Volodymyr Vovchenko, Oleh Savchuk, Roman~V. Poberezhnyuk, Mark~I. Gorenstein,
  and Volker Koch.
\newblock {Connecting fluctuation measurements in heavy-ion collisions with the
  grand-canonical susceptibilities}.
\newblock {\em Phys. Lett. B}, 811:135868, 2020.

\bibitem{koch2025exploringqcdphasediagram}
Volker Koch and Volodymyr Vovchenko.
\newblock Exploring the qcd phase diagram through correlations and
  fluctuations, 2025.

\bibitem{Kitazawa:2012at}
Masakiyo Kitazawa and Masayuki Asakawa.
\newblock {Relation between baryon number fluctuations and experimentally
  observed proton number fluctuations in relativistic heavy ion collisions}.
\newblock {\em Phys.Rev.}, C86:024904, 2012.

\bibitem{Bellwied:2019pxh}
Rene Bellwied, Szabolcs Borsanyi, Zoltan Fodor, Jana~N. Guenther, Jacquelyn
  Noronha-Hostler, Paolo Parotto, Attila Pasztor, Claudia Ratti, and Jamie~M.
  Stafford.
\newblock {Off-diagonal correlators of conserved charges from lattice QCD and
  experiment}.
\newblock {\em Physical Review D}, 101(3), Feb 2020.

\bibitem{Blaizot:2001vr}
J.P. Blaizot, Edmond Iancu, and A.~Rebhan.
\newblock {Quark number susceptibilities from HTL resummed thermodynamics}.
\newblock {\em Phys.Lett.}, B523:143--150, 2001.

\bibitem{Andersen:2012wr}
Jens~O. Andersen, Sylvain Mogliacci, Nan Su, and Aleksi Vuorinen.
\newblock {Quark number susceptibilities from resummed perturbation theory}.
\newblock {\em Phys.Rev.}, D87:074003, 2013.

\bibitem{Haque:2013sja}
Najmul Haque, Jens~O. Andersen, Munshi~G. Mustafa, Michael Strickland, and Nan
  Su.
\newblock {Three-loop HTLpt Pressure and Susceptibilities at Finite Temperature
  and Density}.
\newblock {\em Phys.Rev.}, D89:061701, 2014.

\bibitem{Haque:2013qta}
Najmul Haque, Munshi~G. Mustafa, and Michael Strickland.
\newblock {Quark Number Susceptibilities from Two-Loop Hard Thermal Loop
  Perturbation Theory}.
\newblock {\em JHEP}, 1307:184, 2013.

\bibitem{Fischer:2026vkc}
Christian~S. Fischer and Jan~M. Pawlowski.
\newblock {Phase structure of strong interaction matter from Functional QCD}.
\newblock 6 2026.

\bibitem{Gao:2020qsj}
Fei Gao and Jan~M. Pawlowski.
\newblock {QCD phase structure from functional methods}.
\newblock {\em Phys. Rev. D}, 102(3):034027, 2020.

\bibitem{Fu:2021oaw}
Wei-jie Fu, Xiaofeng Luo, Jan~M. Pawlowski, Fabian Rennecke, Rui Wen, and Shi
  Yin.
\newblock {Hyper-order baryon number fluctuations at finite temperature and
  density}.
\newblock {\em Phys. Rev. D}, 104(9):094047, 2021.

\bibitem{Fu:2023lcm}
Wei-jie Fu, Xiaofeng Luo, Jan~M. Pawlowski, Fabian Rennecke, and Shi Yin.
\newblock {Ripples of the QCD critical point}.
\newblock {\em Phys. Rev. D}, 111(3):L031502, 2025.

\bibitem{Isserstedt:2019pgx}
Philipp Isserstedt, Michael Buballa, Christian~S. Fischer, and Pascal~J.
  Gunkel.
\newblock {Baryon number fluctuations in the QCD phase diagram from
  Dyson-Schwinger equations}.
\newblock {\em Phys. Rev.}, D100(7):074011, 2019.

\bibitem{Bernhardt:2022mnx}
Julian Bernhardt, Christian~S. Fischer, and Philipp Isserstedt.
\newblock {Finite-volume effects in baryon number fluctuations around the QCD
  critical endpoint}.
\newblock {\em Phys. Lett. B}, 841:137908, 2023.

\bibitem{Lu:2025cls}
Yi~Lu, Fei Gao, Yu-xin Liu, and Jan~M. Pawlowski.
\newblock {Finite density signatures of confining and chiral dynamics in QCD
  thermodynamics and fluctuations of conserved charges}.
\newblock 4 2025.

\bibitem{Bazavov:2013dta}
A.~Bazavov, H.~T. Ding, P.~Hegde, O.~Kaczmarek, F.~Karsch, et~al.
\newblock {Strangeness at high temperatures: from hadrons to quarks}.
\newblock {\em Phys. Rev. Lett. 111,}, 082301:082301, 2013.

\bibitem{Vovchenko:2017xad}
Volodymyr Vovchenko, Attila Pasztor, Zoltan Fodor, Sandor~D. Katz, and Horst
  Stoecker.
\newblock {Repulsive baryonic interactions and lattice QCD observables at
  imaginary chemical potential}.
\newblock {\em Phys. Lett.}, B775:71--78, 2017.

\bibitem{Alba:2017mqu}
Paolo Alba et~al.
\newblock {Constraining the hadronic spectrum through QCD thermodynamics on the
  lattice}.
\newblock {\em Phys. Rev.}, D96(3):034517, 2017.

\bibitem{Vovchenko:2019pjl}
Volodymyr Vovchenko and Horst Stoecker.
\newblock {Thermal-FIST: A package for heavy-ion collisions and hadronic
  equation of state}.
\newblock {\em Comput. Phys. Commun.}, 244:295--310, 2019.

\bibitem{Vovchenko:2020lju}
Volodymyr Vovchenko.
\newblock {Hadron resonance gas with van der Waals interactions}.
\newblock {\em Int. J. Mod. Phys. E}, 29(05):2040002, 2020.

\bibitem{Bonati:2015bha}
Claudio Bonati, Massimo D'Elia, Marco Mariti, Michele Mesiti, Francesco Negro,
  and Francesco Sanfilippo.
\newblock {Curvature of the chiral pseudocritical line in QCD: Continuum
  extrapolated results}.
\newblock {\em Phys. Rev.}, D92(5):054503, 2015.

\bibitem{Borsanyi:2013bia}
Szabolcs Borsanyi, Zoltan Fodor, Christian Hoelbling, Sandor~D. Katz, Stefan
  Krieg, et~al.
\newblock {Full result for the QCD equation of state with 2+1 flavors}.
\newblock {\em Phys.Lett.}, B730:99--104, 2014.

\bibitem{HotQCD:2014kol}
A.~Bazavov et~al.
\newblock {Equation of state in ( 2+1 )-flavor QCD}.
\newblock {\em Phys. Rev. D}, 90:094503, 2014.

\bibitem{Borsanyi:2011sw}
Szabolcs Borsanyi, Zoltan Fodor, Sandor~D. Katz, Stefan Krieg, Claudia Ratti,
  et~al.
\newblock {Fluctuations of conserved charges at finite temperature from lattice
  QCD}.
\newblock {\em JHEP}, 1201:138, 2012.

\bibitem{Bazavov:2012jq}
A.~Bazavov et~al.
\newblock {Fluctuations and Correlations of net baryon number, electric charge,
  and strangeness: A comparison of lattice QCD results with the hadron
  resonance gas model}.
\newblock {\em Phys.Rev.}, D86:034509, 2012.

\bibitem{Bellwied:2015lba}
R.~Bellwied, S.~Borsanyi, Z.~Fodor, S.~D. Katz, A.~Pasztor, C.~Ratti, and K.~K.
  Szabo.
\newblock {Fluctuations and correlations in high temperature QCD}.
\newblock {\em Phys. Rev.}, D92(11):114505, 2015.

\bibitem{Bollweg:2021vqf}
D.~Bollweg, J.~Goswami, O.~Kaczmarek, F.~Karsch, Swagato Mukherjee,
  P.~Petreczky, C.~Schmidt, and P.~Scior.
\newblock {Second order cumulants of conserved charge fluctuations revisited:
  Vanishing chemical potentials}.
\newblock {\em Phys. Rev. D}, 104(7), 2021.

\bibitem{Goswami:2026hit}
Jishnu Goswami, Yasumichi Aoki, Hidenori Fukaya, Shoji Hashimoto, Issaku
  Kanamori, Takashi Kaneko, Yoshifumi Nakamura, David Ward, and Yu~Zhang.
\newblock {Quark Number Susceptibilities and Conserved Charge Fluctuations in
  $(2+1)$-flavor QCD with M{\"o}bius domain-wall fermions (MDWF)}.
\newblock 4 2026.

\bibitem{Bollweg:2024epj}
D.~Bollweg, H.~T. Ding, J.~Goswami, F.~Karsch, Swagato Mukherjee, P.~Petreczky,
  and C.~Schmidt.
\newblock {Strangeness-correlations on the pseudocritical line in (2+1)-flavor
  QCD}.
\newblock {\em Phys. Rev. D}, 110(5):054519, 2024.

\bibitem{Note1}
The continuum extrapolation of $\chi _4^Q$ at $T=130$~MeV in Ref.~\cite
  {Bellwied:2015lba} used lattices with up to $N_\tau =32$. However, the
  $N_\tau =24$ and $N_\tau =32$ lattices had smaller volumes, which affected
  the continuum limit. Hence, we do not show this result in this work's plots.

\bibitem{Aarts:2014nba}
Gert Aarts, Chris Allton, Alessandro Amato, Pietro Giudice, Simon Hands, et~al.
\newblock {Electrical conductivity and charge diffusion in thermal QCD from the
  lattice}.
\newblock {\em JHEP}, 1502:186, 2015.

\bibitem{Borsanyi:2023wno}
Szabolcs Borsanyi, Zoltan Fodor, Jana~N. Guenther, Sandor~D. Katz, Paolo
  Parotto, Attila Pasztor, David Pesznyak, Kalman~K. Szabo, and Chik~Him Wong.
\newblock {Continuum-extrapolated high-order baryon fluctuations}.
\newblock {\em Phys. Rev. D}, 110(1):L011501, 2024.

\bibitem{Dashen:1969ep}
Roger Dashen, Shang-Keng Ma, and Herbert~J. Bernstein.
\newblock {S Matrix formulation of statistical mechanics}.
\newblock {\em Phys.Rev.}, 187:345--370, 1969.

\bibitem{Adamczyk:2014fia}
L.~Adamczyk et~al.
\newblock {Beam energy dependence of moments of the net-charge multiplicity
  distributions in Au+Au collisions at RHIC}.
\newblock {\em Phys.Rev.Lett.}, 113:092301, 2014.

\bibitem{Adare:2015aqk}
A.~Adare et~al.
\newblock {Measurement of higher cumulants of net-charge multiplicity
  distributions in Au$+$Au collisions at $\sqrt{s_{_{NN}}}=7.7-200$ GeV}.
\newblock 2015.

\bibitem{Takaishi:1996xj}
Tetsuya Takaishi.
\newblock {Heavy quark potential and effective actions on blocked
  configurations}.
\newblock {\em Phys.Rev.}, D54:1050--1053, 1996.

\bibitem{Capitani:2006ni}
Stefano Capitani, Stephan Durr, and Christian Hoelbling.
\newblock {Rationale for UV-filtered clover fermions}.
\newblock {\em JHEP}, 11:028, 2006.

\bibitem{BMW:2012hcm}
Szabolcs Borsanyi et~al.
\newblock {High-precision scale setting in lattice QCD}.
\newblock {\em JHEP}, 09:010, 2012.

\bibitem{Borsanyi:2024xrx}
Szabolcs Borsanyi, Zoltan Fodor, Jana~N. Guenther, Paolo Parotto, Attila
  Pasztor, Ludovica Pirelli, Kalman~K. Szabo, and Chik~Him Wong.
\newblock {QCD deconfinement transition line up to \ensuremath{\mu}B=400\,\,MeV
  from finite volume lattice simulations}.
\newblock {\em Phys. Rev. D}, 110(11):114507, 2024.

\bibitem{Allton:2002zi}
C.R. Allton, S.~Ejiri, S.J. Hands, O.~Kaczmarek, F.~Karsch, et~al.
\newblock {The QCD thermal phase transition in the presence of a small chemical
  potential}.
\newblock {\em Phys.Rev.}, D66:074507, 2002.

\bibitem{Borsanyi:2020mff}
Sz. Borsanyi et~al.
\newblock {Leading hadronic contribution to the muon magnetic moment from
  lattice QCD}.
\newblock {\em Nature}, 593(7857):51--55, 2021.

\bibitem{Hasenfratz:1991ax}
A.~Hasenfratz and D.~Toussaint.
\newblock {Canonical ensembles and nonzero density quantum chromodynamics}.
\newblock {\em Nucl. Phys.}, B371:539--549, 1992.

\bibitem{Bellwied:2015rza}
R.~Bellwied, S.~Borsanyi, Z.~Fodor, J.~G{\"u}nther, S.~D. Katz, C.~Ratti, and
  K.~K. Szabo.
\newblock {The QCD phase diagram from analytic continuation}.
\newblock {\em Phys. Lett.}, B751:559--564, 2015.

\bibitem{Abuali:2025tbd}
Ahmed Abuali, Szabolcs Bors{\'a}nyi, Zolt{\'a}n Fodor, Johannes Jahan, Micheal
  Kahangirwe, Paolo Parotto, Attila P{\'a}sztor, Claudia Ratti, Hitansh Shah,
  and Seth~A. Trabulsi.
\newblock {New 4D lattice QCD equation of state: Extended density coverage from
  a generalized T' expansion}.
\newblock {\em Phys. Rev. D}, 112(5):054502, 2025.

\bibitem{SanMartin:2023zhv}
Jordi~Salinas San~Martin, Renan Hirayama, Jan Hammelmann, Jamie~M. Karthein,
  Paolo Parotto, Jacquelyn Noronha-Hostler, Claudia Ratti, and Hannah Elfner.
\newblock {Thermodynamics of an updated hadronic resonance list and influence
  on hadronic transport}.
\newblock 9 2023.

\bibitem{Dashen:1969eg}
Roger~F. Dashen.
\newblock {Chiral SU(3) x SU(3) as a symmetry of the strong interactions}.
\newblock {\em Phys. Rev.}, 183:1245--1260, 1969.

\bibitem{Lo:2017lym}
Pok~Man Lo, Bengt Friman, Krzysztof Redlich, and Chihiro Sasaki.
\newblock {S-matrix analysis of the baryon electric charge correlation}.
\newblock {\em Phys. Lett. B}, 778:454--458, 2018.

\bibitem{Andronic:2018qqt}
Anton Andronic, Peter Braun-Munzinger, Bengt Friman, Pok~Man Lo, Krzysztof
  Redlich, and Johanna Stachel.
\newblock {The thermal proton yield anomaly in Pb-Pb collisions at the LHC and
  its resolution}.
\newblock {\em Phys. Lett.}, B792:304--309, 2019.

\bibitem{Honek:2026exd}
Vojtech Honek, Pok~Man Lo, and Boris Tomasik.
\newblock {$S$-matrix calculation of $BQ$ correlation at finite baryon
  density}.
\newblock 4 2026.

\bibitem{Garcia-Martin:2011iqs}
R.~Garcia-Martin, R.~Kaminski, J.~R. Pelaez, J.~Ruiz~de Elvira, and F.~J.
  Yndurain.
\newblock {The Pion-pion scattering amplitude. IV: Improved analysis with once
  subtracted Roy-like equations up to 1100 MeV}.
\newblock {\em Phys. Rev. D}, 83:074004, 2011.

\bibitem{Pelaez:2016klv}
J.~R. Pel{\'a}ez, A.~Rodas, and J.~Ruiz~de Elvira.
\newblock {Strange resonance poles from $K\pi $ scattering below 1.8 GeV}.
\newblock {\em Eur. Phys. J. C}, 77(2):91, 2017.

\bibitem{Venugopalan:1992hy}
R.~Venugopalan and M.~Prakash.
\newblock {Thermal properties of interacting hadrons}.
\newblock {\em Nucl.Phys.}, A546:718--760, 1992.

\bibitem{Broniowski:2015oha}
Wojciech Broniowski, Francesco Giacosa, and Viktor Begun.
\newblock {Cancellation of the $\sigma$ meson in thermal models}.
\newblock {\em Phys. Rev. C}, 92(3):034905, 2015.

\bibitem{Note2}
We note that a similar treatment of light meson interactions in the S-matrix
  HRG model, with similar results, was presented at the YSTAR2016
  conference~\protect \url
  {https://www.jlab.org/conferences/YSTAR2016/talks/thursday/huovinen.pdf}, but
  we are not aware of follow-up published results.

\bibitem{Vovchenko:2018fmh}
Volodymyr Vovchenko, Mark~I. Gorenstein, and Horst Stoecker.
\newblock {Finite resonance widths influence the thermal-model description of
  hadron yields}.
\newblock {\em Phys. Rev. C}, 98(3):034906, 2018.

\bibitem{Savchuk:2020yxc}
Oleh Savchuk, Yehor Bondar, Oleksandr Stashko, Roman~V. Poberezhnyuk, Volodymyr
  Vovchenko, Mark~I. Gorenstein, and Horst Stoecker.
\newblock {Bose-Einstein condensation phenomenology in systems with repulsive
  interactions}.
\newblock {\em Phys. Rev. C}, 102(3):035202, 2020.

\bibitem{Vovchenko:2020crk}
Volodymyr Vovchenko, Bastian~B. Brandt, Francesca Cuteri, Gergely
  Endr{\H{o}}di, Fazlollah Hajkarim, and J{\"u}rgen Schaffner-Bielich.
\newblock {Pion Condensation in the Early Universe at Nonvanishing Lepton
  Flavor Asymmetry and Its Gravitational Wave Signatures}.
\newblock {\em Phys. Rev. Lett.}, 126(1):012701, 2021.

\bibitem{Lowdon:2022xcl}
Peter Lowdon and Owe Philipsen.
\newblock {Pion spectral properties above the chiral crossover of QCD}.
\newblock {\em JHEP}, 10:161, 2022.

\bibitem{Vovchenko:2024pvk}
Volodymyr Vovchenko.
\newblock {Density correlations under global and local charge conservation}.
\newblock {\em Phys. Rev. C}, 110(6):L061902, 2024.

\bibitem{Son:2000xc}
D.~T. Son and Misha~A. Stephanov.
\newblock {QCD at finite isospin density}.
\newblock {\em Phys. Rev. Lett.}, 86:592--595, 2001.

\bibitem{Borsanyi:2021sxv}
S.~Bors\'anyi, Z.~Fodor, J.~N. Guenther, R.~Kara, S.~D. Katz, P.~Parotto,
  A.~P\'asztor, C.~Ratti, and K.~K. Szab\'o.
\newblock {Lattice QCD equation of state at finite chemical potential from an
  alternative expansion scheme}.
\newblock {\em Phys. Rev. Lett.}, 126(23):232001, 2021.

\end{thebibliography}

\end{document}